# Score predictor factor analysis:
# Reproducing observed covariances by means of factor score predictors

André Beauducel & Norbert Hilger

University of Bonn, Institute of Psychology

Address for correspondence:

Dr. A. Beauducel

Institute of Psychology

Rheinische Friedrich-Wilhelms-Universität Bonn

Kaiser-Karl-Ring 9

53111 Bonn

beauducel@uni-bonn.de

**Abstract**

The model implied by factor score predictors does not reproduce the non-diagonal elements of the observed covariance matrix as well as the factor loadings. It is therefore investigated whether it is possible to estimate factor loadings for which the model implied by the factor score predictors optimally reproduces the non-diagonal elements of the observed covariance matrix. Accordingly, loading estimates are proposed for which the model implied by the factor score predictors allows for a least-squares approximation of the non-diagonal elements of the observed covariance matrix. This estimation method is termed Score predictor factor analysis and algebraically compared with Minres factor analysis as well as principal component analysis. A population based and a sample based simulation study was performed in order to compare Score predictor factor analysis, Minres factor analysis, and principal component analysis. It turns out that the non-diagonal elements of the observed covariance matrix can more exactly be reproduced from the factor score predictors computed from Score predictor factor analysis than from the factor score predictors computed from Minres factor analysis and from principal components. Moreover, Score predictor factor analysis can be helpful to identify factors when the factor model does not perfectly fit to the data because of model error.

The factor model reproduces the observed covariances from the loadings and inter-correlations of the common factors as well as from the loadings of the unique factors. The factor loadings and factor inter-correlations are typically identified by means of rotational criteria in exploratory factor analysis or by means of model specification in the context of confirmatory factor analysis. However, the individual scores on the factors are indeterminate (Grice, 2001; Guttman, 1955; Wilson, 1929) even when all parameters (loadings, inter-factor correlations, error terms) of the factor model are identified. Since the individual scores on the factors themselves are indeterminate, individual scores on so-called 'factor score estimators' (McDonald, 1981), sometimes called 'factor score predictors' (Krijnen, 2006), are computed when individual scores on the common factors are needed. The need for individual scores can occur in different areas, for example, in the context of psychological assessment, when individuals are selected for a job or when an optimal treatment has to be assigned to an individual.

Since 'estimating' scores that are not uniquely defined may be regarded as unconventional (Schönemann & Steiger, 1976), Krijnen's (2006) term 'factor score predictor' will be used in the following. However, even this term does not describe that, in fact, scores are only constructed (McDonald & Burr, 1967) in a way that they reflect some aspects of the original, but indeterminate factors. Since the factor score predictors are not identical with the factors themselves, three criteria for the evaluation of factor score predictors are typically considered (Grice, 2001): The maximal correlation of the factor score predictor with the factor (validity), that factor score predictors do not correlate with non-corresponding factors (univocality), and the similarity of the inter-correlations of the factor score predictors with the inter-correlations of the factors (correlational accuracy). The importance of an evaluation of the quality of factor score predictors has been acknowledged repeatedly (Ferrando & Lorenzo-Seva, in press).

Although the aim of the factor model is to reproduce the observed covariances, the covariances that are reproduced from factor score predictors are not the same as the covariances that are reproduced from the factors themselves (Beauducel, 2007). Even when the factor model reproduces the observed covariances quite well, the factor score predictors typically will not reproduce the observed covariances as well. Nevertheless, factor score predictors that optimally represent the factors should reproduce the covariances as well as the factors themselves. Accordingly, Beauducel and Hilger (2015) proposed to evaluate how well factor score predictors can reproduce the observed covariance matrix as a further criterion for the evaluation of factor score predictors. They noted that especially the reproduction of the non-diagonal elements of the observed covariance matrix by means of the common factors, as performed with Minres factor analysis (MFA, Comrey, 1962; Harman & Jones, 1966; Harman, 1976), represents this aspect of the factor model. Moreover, Beauducel and Hilger (2015) found that an optimal reproduction of the non-diagonal elements of the observed covariance matrix by the factor score predictors is possible when there is a single variable with a perfect loading on each factor. However, perfect factor loadings rarely occur in empirical research. Therefore, the weighting of observed variables resulting in factor score predictors that optimally reproduce the non-diagonal elements of the observed covariance matrix is widely unknown. In order to close this gap it is proposed to estimate factor loadings in a way that not the loadings but the factor score predictors computed from the loadings optimally reproduce the non-diagonal elements of the observed covariance matrix.

In sum, (1) factor score predictors do not reproduce the non-diagonal elements of the observed covariance matrix as well as the common factors do. (2) The reproduction of the non-diagonal elements of the observed covariance can be considerably improved when a single variable with a perfect loading occurs on each common factor. This leads to the question whether it is possible to find a factor model for which the corresponding factor score predictors optimally reproduce the non-diagonal elements of the observed covariance matrix. Accordingly, the present paper is devoted to the investigation of a least-squares approximation of the non-diagonal elements of the covariance matrix reproduced from the factor score predictors to the non-diagonal elements of the observed covariance matrix. After some definitions, a model is proposed for which the covariance matrix reproduced from the model implied by factor score predictors allows for a least-squares approximation of the observed covariances. This is a specific estimation method in the context of the factor model because loadings are estimated for which the factor score predictors that are computed from the loadings optimally reproduce the non-diagonal elements of the observed covariance matrix. Therefore, the method is termed Score-predictor factor analysis and compared with Minres factor analysis as well as principal components analysis by means of algebraic considerations, a small population based simulation, and a sample based simulation study.

### Definitions

Let **x** be a random vector of order $p$ representing a population of observed scores with E(**x**)=0. According to the factor model **x** can be decomposed by

$$\mathbf{x} = \mathbf{\Lambda f} + \mathbf{u}, \tag{1}$$

where **Λ** is a $p \times q$ loading matrix, **f** is a random vector of order $q$ representing the common factors with E(**f**)=0, E(**ff'**)=**Φ**, and $diag(\mathbf{\Phi}) = \mathbf{I}$, and **u** is a random vector of order $p$ representing the unique variance of each variable, with E(**u**) = 0 and E(**uu'**) = $diag(E(\mathbf{uu'}))$ = **Ψ**². Accordingly, the covariance matrix of observed variables that is reproduced from the factor model can be written as

$$E(\mathbf{xx'}) = \mathbf{\Sigma} = \mathbf{\Lambda \Phi \Lambda'} + \mathbf{\Psi}^2. \tag{2}$$

In contrast, the covariances that are reproduced from linear combinations of observed variables are given by

$$\mathbf{\Sigma_r} = \mathbf{\Sigma B}(\mathbf{B'\Sigma B})^{-1}\mathbf{B'\Sigma}, \tag{3}$$

where **B** represents a matrix of weights for the observed variables. Factor score predictors are linear combinations of the observed variables. Beauducel (2007) has shown for the regression factor score predictor (Thurstone, 1935), the Bartlett factor score predictor (Bartlett, 1937), and the McDonald factor score predictor (McDonald, 1981) that Equation 3 can be expressed in terms of the factor loadings, with

$$\mathbf{\Sigma_r} = \mathbf{\Lambda}(\mathbf{\Lambda'\Sigma}^{-1}\mathbf{\Lambda})^{-1}\mathbf{\Lambda'}. \tag{4}$$

Thus, $\mathbf{\Sigma_r}$ are the covariances that are reproduced from the model that is implied by the abovementioned factor score predictors. Since the model implied by the factor score predictors depends on **Λ**, it is proposed to find loading estimates, for which the non-diagonal elements of $\mathbf{\Sigma_r}$ are a least squares approximation of the non-diagonal elements of **Σ**.

**Approximation of the non-diagonal elements of the observed sample covariance matrix**

As factor model parameters will typically be estimated on the basis of covariance matrices observed in the sample, the estimation procedures are given for the samples. For the sample, Equation 2 can be written as

$$\mathbf{S} \approx \hat{\mathbf{\Lambda}}\hat{\mathbf{\Phi}}\hat{\mathbf{\Lambda}}' + \hat{\mathbf{\Psi}}^2, \qquad (5)$$

and Equation 4 can be written as

$$\hat{\mathbf{\Sigma}}_\mathbf{r} = \hat{\mathbf{\Lambda}}(\hat{\mathbf{\Lambda}}'\mathbf{S}^{-1}\hat{\mathbf{\Lambda}})^{-1}\hat{\mathbf{\Lambda}}', \qquad (6)$$

where $\mathbf{S}$ is the sample covariance matrix. The condition for MFA can be expressed as

$$tr\left[(\mathbf{S} - \hat{\mathbf{\Lambda}}_M\hat{\mathbf{\Lambda}}'_M - diag(\mathbf{S} - \hat{\mathbf{\Lambda}}_M\hat{\mathbf{\Lambda}}'_M))'(\mathbf{S} - \hat{\mathbf{\Lambda}}_M\hat{\mathbf{\Lambda}}'_M - diag(\mathbf{S} - \hat{\mathbf{\Lambda}}_M\hat{\mathbf{\Lambda}}'_M))\right] = \min, \qquad (7)$$

where $\hat{\mathbf{\Lambda}}_M$ is the loading matrix of MFA. According to Harman and Jones (1966), the following classical principal factor procedure allows for finding $\hat{\mathbf{\Lambda}}_M$:

1) Start with an arbitrary $p \times p$ diagonal matrix $\mathbf{H}$ and compute $\mathbf{S} - diag(\mathbf{S}) + \mathbf{H}$.

2) Perform the eigen-decomposition $\mathbf{S} - diag(\mathbf{S}) + \mathbf{H} = \mathbf{K}\mathbf{\Gamma}\mathbf{K}'$, with eigenvectors $\mathbf{K}$ and with $\mathbf{\Gamma}$ containing the eigenvalues in descending order.

3) Determine the $q$ common factors and compute $\hat{\mathbf{\Lambda}}_M = \mathbf{K}_q\mathbf{\Gamma}_q^{1/2}$, where $\mathbf{K}_q$ is a $p \times q$ submatrix of $\mathbf{K}$ and $\mathbf{\Gamma}_q$ is the $q \times q$ submatrix of $\mathbf{\Gamma}$.

4) Determine the reproduced communalities by means of $\mathbf{H} = diag(\hat{\mathbf{\Lambda}}_M\hat{\mathbf{\Lambda}}'_M)$.

5) Insert $\hat{\mathbf{\Lambda}}_M$ into Equation 7 and check whether a convergence criterion is met. The convergen ce criterion is defined by the difference between the previous value resulting from Equation 7 and the current values resulting from Equation 7.

6) Repeat step 2-5 until the convergence criterion is met.

By means of these steps MFA estimates the loadings of orthogonal factors that are conform to the condition expressed in Equation 7. These loading estimates optimally reproduce the non-diagonal elements of the observed sample covariance matrix. The present approach is to replace $\hat{\mathbf{\Lambda}}_M\hat{\mathbf{\Lambda}}'_M$ in Equation 7 by the covariances reproduced from the factor score predictor (Equation 6). This leads to

$$tr[(\mathbf{S}-\hat{\mathbf{\Lambda}}_{os}(\hat{\mathbf{\Lambda}}'_{os}\mathbf{S}^{-1}\hat{\mathbf{\Lambda}}_{os})^{-1}\hat{\mathbf{\Lambda}}'_{os} - diag(\mathbf{S}-\hat{\mathbf{\Lambda}}_{os}(\hat{\mathbf{\Lambda}}'_{os}\mathbf{S}^{-1}\hat{\mathbf{\Lambda}}_{os})^{-1}\hat{\mathbf{\Lambda}}'_{os}))' \\ (\mathbf{S}-\hat{\mathbf{\Lambda}}_{os}(\hat{\mathbf{\Lambda}}'_{os}\mathbf{S}^{-1}\hat{\mathbf{\Lambda}}_{os})^{-1}\hat{\mathbf{\Lambda}}'_{os} - diag(\mathbf{S}-\hat{\mathbf{\Lambda}}_{os}(\hat{\mathbf{\Lambda}}'_{os}\mathbf{S}^{-1}\hat{\mathbf{\Lambda}}_{os})^{-1}\hat{\mathbf{\Lambda}}'_{os}))] = \min, \quad (8)$$

where $\hat{\mathbf{\Lambda}}_{os}$ is a loading pattern of the estimation method proposed here, which is called Score predictor factor analysis (SPFA). The covariance of the SPFA factors is $(\hat{\mathbf{\Lambda}}'_{os}\mathbf{S}^{-1}\hat{\mathbf{\Lambda}}_{os})^{-1}$. According to this condition, a loadings are estimated for which the non-diagonal elements of the covariance matrix reproduced from the abovementioned factor score predictors (regression, Bartlett, McDonald) are a least squares approximation of the non-diagonal elements of the observed covariance matrix. The corresponding orthogonal loading pattern is

$$\hat{\mathbf{\Lambda}}_s = \hat{\mathbf{\Lambda}}_{os}(\hat{\mathbf{\Lambda}}'_{os}\mathbf{S}^{-1}\hat{\mathbf{\Lambda}}_{os})^{-1/2}, \quad (9)$$

where "-1/2" denotes the inverse of the symmetric square-root. The corresponding unique variance is $\hat{\mathbf{\Psi}}_s^2 = diag(\mathbf{S}-\hat{\mathbf{\Lambda}}_s\hat{\mathbf{\Lambda}}'_s)$. The estimation of SPFA loadings can be performed by means of the same algorithm as for MFA. The only difference is that

$$\hat{\mathbf{\Lambda}}_s = \hat{\mathbf{\Lambda}}_{os}(\hat{\mathbf{\Lambda}}'_{os}\mathbf{S}^{-1}\hat{\mathbf{\Lambda}}_{os})^{-1/2} = \mathbf{K}_q\mathbf{\Gamma}_q^{1/2}(\mathbf{\Gamma}_q^{1/2}\mathbf{K}'_q\mathbf{S}^{-1}\mathbf{K}_q\mathbf{\Gamma}_q^{1/2})^{-1/2} = \mathbf{K}_q(\mathbf{K}'_q\mathbf{S}^{-1}\mathbf{K}_q)^{-1/2} \quad (10)$$

is inserted instead of $\hat{\mathbf{\Lambda}}_M$ in step 3 and the following steps of the abovementioned algorithm.

**Reproducing covariances from loadings and from the corresponding factor score predictor models**

The MFA and SPFA estimation procedures were described for the sample observed covariance matrix because these procedures will typically be calculated for these matrices. However, in the following, the population observed covariance matrices will be used in order to describe further properties of SPFA and MFA.

Equations 2 and 4 of the factor model imply $\mathbf{\Sigma} - diag(\mathbf{\Sigma}) \neq \mathbf{\Sigma_r} - diag(\mathbf{\Sigma_r})$ for $\mathbf{\Lambda}'\mathbf{\Sigma}^{-1}\mathbf{\Lambda} \neq \mathbf{I}$. It follows from Beauducel and Hilger (2015, Eq. 7) that

$$\mathbf{\Lambda}(\mathbf{\Lambda}'\mathbf{\Sigma}^{-1}\mathbf{\Lambda})^{-1}\mathbf{\Lambda}' = \mathbf{\Lambda}\mathbf{\Lambda}' + \mathbf{\Lambda}(\mathbf{\Lambda}'\mathbf{\Psi}^{-2}\mathbf{\Lambda})^{-1}\mathbf{\Lambda}', \quad (11)$$

so that $\mathbf{\Lambda}(\mathbf{\Lambda}'\mathbf{\Sigma}^{-1}\mathbf{\Lambda})^{-1}\mathbf{\Lambda}' = \mathbf{\Lambda}\mathbf{\Lambda}'$ if $(\mathbf{\Lambda}'\mathbf{\Psi}^{-2}\mathbf{\Lambda})^{-1} = 0$. If one loading per column of $\mathbf{\Lambda}$ approaches one, the corresponding element in $\mathbf{\Psi}$ approaches zero, so that $(\mathbf{\Lambda}'\mathbf{\Psi}^{-2}\mathbf{\Lambda})^{-1}$ also approaches zero (Beauducel & Hilger, 2015). Accordingly, $\mathbf{\Sigma} - diag(\mathbf{\Sigma})$ will become similar to $\mathbf{\Sigma_r} - diag(\mathbf{\Sigma_r})$ under this condition. This implies that the corresponding elements of the covariance matrices reproduced from MFA will also become similar when one loading per column of $\mathbf{\Lambda}$ approaches one.

For SPFA, the non-diagonal elements of covariance matrices reproduced from the loadings are given by $\Sigma_s - diag(\Sigma_s) = \Lambda_s\Lambda_s' - diag(\Lambda_s\Lambda_s')$. The non-diagonal elements of the covariance matrix reproduced from the factor score predictors are given by $\Sigma_{rs} - diag(\Sigma_{rs}) = \Lambda_s(\Lambda_s'\Sigma^{-1}\Lambda_s)^{-1}\Lambda_s' - diag(\Lambda_s(\Lambda_s'\Sigma^{-1}\Lambda_s)^{-1}\Lambda_s')$. Theorem 1 describes that $\Sigma_s - diag(\Sigma_s) = \Sigma_{rs} - diag(\Sigma_{rs})$ always holds for SPFA.

**Theorem 1.** $\Sigma_s - diag(\Sigma_s) = \Sigma_{rs} - diag(\Sigma_{rs})$.

*Proof.* Writing Equation 9 for the population yields $\Lambda_s = \Lambda_{os}(\Lambda_{os}'\Sigma^{-1}\Lambda_{os})^{-1/2}$, which implies

$$\begin{aligned}\Lambda_s\Lambda_s' &= \Lambda_{os}(\Lambda_{os}'\Sigma^{-1}\Lambda_{os})^{-1}\Lambda_{os}' \\ &= \Lambda_{os}(\Lambda_{os}'\Sigma^{-1}\Lambda_{os})^{-1/2}((\Lambda_{os}'\Sigma^{-1}\Lambda_{os})^{-1/2}\Lambda_{os}'\Sigma^{-1}\Lambda_{os}(\Lambda_{os}'\Sigma^{-1}\Lambda_{os})^{-1/2})^{-1}(\Lambda_{os}'\Sigma^{-1}\Lambda_{os})^{-1/2}\Lambda_{os}' \quad (12)\\ &= \Lambda_s(\Lambda_s'\Sigma^{-1}\Lambda_s)^{-1}\Lambda_s'.\end{aligned}$$

This completes the proof. □

Theorem 1 implies that the non-diagonal elements of the covariance matrix reproduced from the orthogonal common factor loadings $\Lambda_s$ are identical to the non-diagonal elements of the covariances reproduced from the regression-, Bartlett-, and McDonald factor score predictors computed from $\Lambda_s$ and $\Psi_s^2$. Thus, the fit of the SPFA model to the non-diagonal elements of the observed covariance matrix is equal to the model fit implied by the SPFA factor score predictors.

Although this is usually not the case for MFA, the non-diagonal elements of $\Sigma_A$, the covariance matrix reproduced from the loadings **A** of principal component analysis (PCA), are identical to the non-diagonal elements of $\Sigma_{rc}$, the covariance matrix reproduced from the orthogonal principal component scores **c**. This follows from $\Sigma_A = \Sigma_{rc}$ which is shown in Theorem 2.

**Theorem 2.** *If* $\mathbf{x} = \mathbf{Ac}, \mathbf{cc}' = \mathbf{I}, $ *and* $\Sigma_A = \mathbf{AA}'$ *then* $\Sigma_A = \Sigma_{rc}$.

*Proof.* The component scores are $(\mathbf{A'A})^{-1}\mathbf{A'x} = \mathbf{c}$ so that the corresponding weights $\mathbf{B}_A = \mathbf{A}(\mathbf{A'A})^{-1}$ can be entered into Equation 3. This yields

$$\Sigma_{rc} = \Sigma\mathbf{A}(\mathbf{A'A})^{-1}((\mathbf{A'A})^{-1}\mathbf{A'}\Sigma\mathbf{A}(\mathbf{A'A})^{-1})^{-1}(\mathbf{A'A})^{-1}\mathbf{A'}\Sigma. \quad (13)$$

It follows from $(\mathbf{A'A})^{-1}\mathbf{A'}\Sigma = \mathbf{cx'}$ and from $\mathbf{xc'} = \mathbf{Acc'} = \mathbf{A}$ that Equation 13 can be written as

$$\Sigma_{rc} = \mathbf{xc'}(\mathbf{cx'}\Sigma^{-1}\mathbf{xc'})^{-1}\mathbf{cx'} = \mathbf{A}(\mathbf{A'}\Sigma^{-1}\mathbf{A})^{-1}\mathbf{A'}. \quad (14)$$

The covariance matrix reproduced from the component loadings is

$$\Sigma_A = AA' = Acc'A' = A(A'A)^{-1}A'\Sigma\Sigma^{-1}\Sigma A(A'A)^{-1}A' \\ = AA'\Sigma^{-1}AA' = AA'\Sigma^{-1}A(A'\Sigma^{-1}A)^{-1}A'\Sigma^{-1}AA'. \quad (15)$$

It follows from $(A'A)^{-1}A' = A'\Sigma^{-1}$ that

$$\Sigma_A = AA'\Sigma^{-1}A(A'\Sigma^{-1}A)^{-1}A'\Sigma^{-1}AA' = A(A'\Sigma^{-1}A)^{-1}A'. \quad (16)$$

This completes the proof. □

To summarize, the PCA loadings and -scores as well as the SPFA loadings and score predictors reproduce the non-diagonal elements of the observed covariance matrix equally well. In contrast, the factor model and MFA has this property only when one element of each column has a perfect communality.

**Model error and the SPFA model**

It follows from Equation 12 that in SPFA the condition $\Lambda'_s\Sigma^{-1}\Lambda_s = I$ is imposed on the loadings. Accordingly, the SPFA factor model is more restrictive than the conventional common factor model. It follows for $\Lambda'\Sigma^{-1}\Lambda \neq I$ that $\Sigma = \Lambda\Lambda' + \Psi^2 \neq \Lambda_s\Lambda'_s + \Psi_s^2$ so that the SPFA model does not hold in all populations where the common factor model holds. In consequence, model error is more likely and more pronounced for SPFA than for MFA. Even in a population without model error of MFA, i.e., for $tr\left[(\Sigma - \Lambda_M\Lambda'_M - diag(\Sigma - \Lambda_M\Lambda'_M))'(\Sigma - \Lambda_M\Lambda'_M - diag(\Sigma - \Lambda_M\Lambda'_M))\right] = 0$, the error of the model implied by the MFA factor score predictors could be substantial, i.e., $tr\left[(\Sigma - \Lambda_M H^{-1}\Lambda'_M - diag(\Sigma - \Lambda_M H^{-1}\Lambda'_M))'(\Sigma - \Lambda_M H^{-1}\Lambda'_M - diag(\Sigma - \Lambda_M H^{-1}\Lambda'_M))\right] > 0$, with $H = \Lambda'_M\Sigma^{-1}\Lambda_M$. In contrast, SPFA finds a least squares approximation of the factor score predictor model to the observed covariances (Equation 8). For the model of MFA and SPFA one might therefore expect

$$\begin{aligned} &tr\left[(\Sigma - \Lambda_M\Lambda'_M - diag(\Sigma - \Lambda_M\Lambda'_M))'(\Sigma - \Lambda_M\Lambda'_M - diag(\Sigma - \Lambda_M\Lambda'_M))\right] \leq \\ &tr\left[(\Sigma - \Lambda_s\Lambda'_s - diag(\Sigma - \Lambda_s\Lambda'_s))'(\Sigma - \Lambda_s\Lambda'_s - diag(\Sigma - \Lambda_s\Lambda'_s))\right] \leq \\ &tr\left[(\Sigma - \Lambda_M H^{-1}\Lambda'_M - diag(\Sigma - \Lambda_M H^{-1}\Lambda'_M))'(\Sigma - \Lambda_M H^{-1}\Lambda'_M - diag(\Sigma - \Lambda_M H^{-1}\Lambda'_M))\right]. \end{aligned} \quad (17)$$

Thus, one would expect that the model error of MFA is smaller or equal to the model error of SPFA and that the model error of the factor score predictor model derived from SPFA is smaller or equal to the model error of the factor score predictor model derived from MFA.

Dividing the traces in Equation 17 by $p(p-1)$ and taking the square root yields the standardized root mean squared residual for non-diagonal elements (SRMR$_{ND}$) as an index of model error that has been used elsewhere (Beauducel & Hilger, in press). In the following, the relationship between the SRMR$_{ND}$ of the factor score predictor model derived from MFA will be compared with the SRMR$_{ND}$ of SPFA and PCA for population data that fit perfectly to the factor model. PCA will be included as a frame of reference because PCA has the same property as SPFA in that the model fit computed from the loadings equals the model fit that is computed from the scores (Theorem 2).

**Population simulation for MFA, SPFA, and PCA for data without factor model error**

The simulations were performed with IBM SPSS Version 24. For $\Lambda'\Sigma^{-1}\Lambda = \Lambda_s'\Sigma^{-1}\Lambda_s = I$ MFA and SPFA should have an identical model fit in terms of Equation 17. According to Equation 11, this condition holds when one variable on each common factor has a perfect loading. Therefore, a population-based simulation study starting with a loading matrix $\Lambda_i$ containing one perfect loading per factor and a set of constant non-zero loadings on each factor was performed. In the next step, the perfect loadings were reduced by a decrement of .01 until all non-zero loadings were nearly equal. In the following example, the initial loading matrix $\Lambda_i$ and the final loading matrix $\Lambda_f$ is given for $q = 3$ factors, $p = 9$ variables, and $p/q = 3$.

$$\Lambda_i = \begin{bmatrix} 1.00 & .00 & .00 \\ .50 & .00 & .00 \\ .50 & .00 & .00 \\ .00 & 1.00 & .00 \\ .00 & .50 & .00 \\ .00 & .50 & .00 \\ .00 & .00 & 1.00 \\ .00 & .00 & .50 \\ .00 & .00 & .50 \end{bmatrix}, \dots, \Lambda_f = \begin{bmatrix} .51 & .00 & .00 \\ .50 & .00 & .00 \\ .50 & .00 & .00 \\ .00 & .51 & .00 \\ .00 & .50 & .00 \\ .00 & .50 & .00 \\ .00 & .00 & .51 \\ .00 & .00 & .50 \\ .00 & .00 & .50 \end{bmatrix}. \quad (18)$$

For $q = 3$, $p = 15$, and $p/q = 5$ similar matrices were generated. Figure 1 A and B illustrates the fit of the model implied by the factor/component scores of MFA, SPFA, and PCA in terms of SRMR$_{ND}$ for these loading matrices. The size of the largest loading on each factor is given on the x-axis and the SRMR$_{ND}$ is given on the y-axis. The models implied by PCA scores have consistently the largest SRMR$_{ND}$, i.e., the lowest fit, whereas the models implied by SPFA factor score predictors have the lowest SRMR$_{ND}$. The SRMR$_{ND}$ of the model implied by the MFA factor score predictors is in between. This shows that the relationship between the MFA loading based SRMR$_{ND}$, the SPFA factor score predictor based SRMR$_{ND}$, and the MFA factor score predictor based SRMR$_{ND}$ in Figure 1 is as predicted in Equation 17.

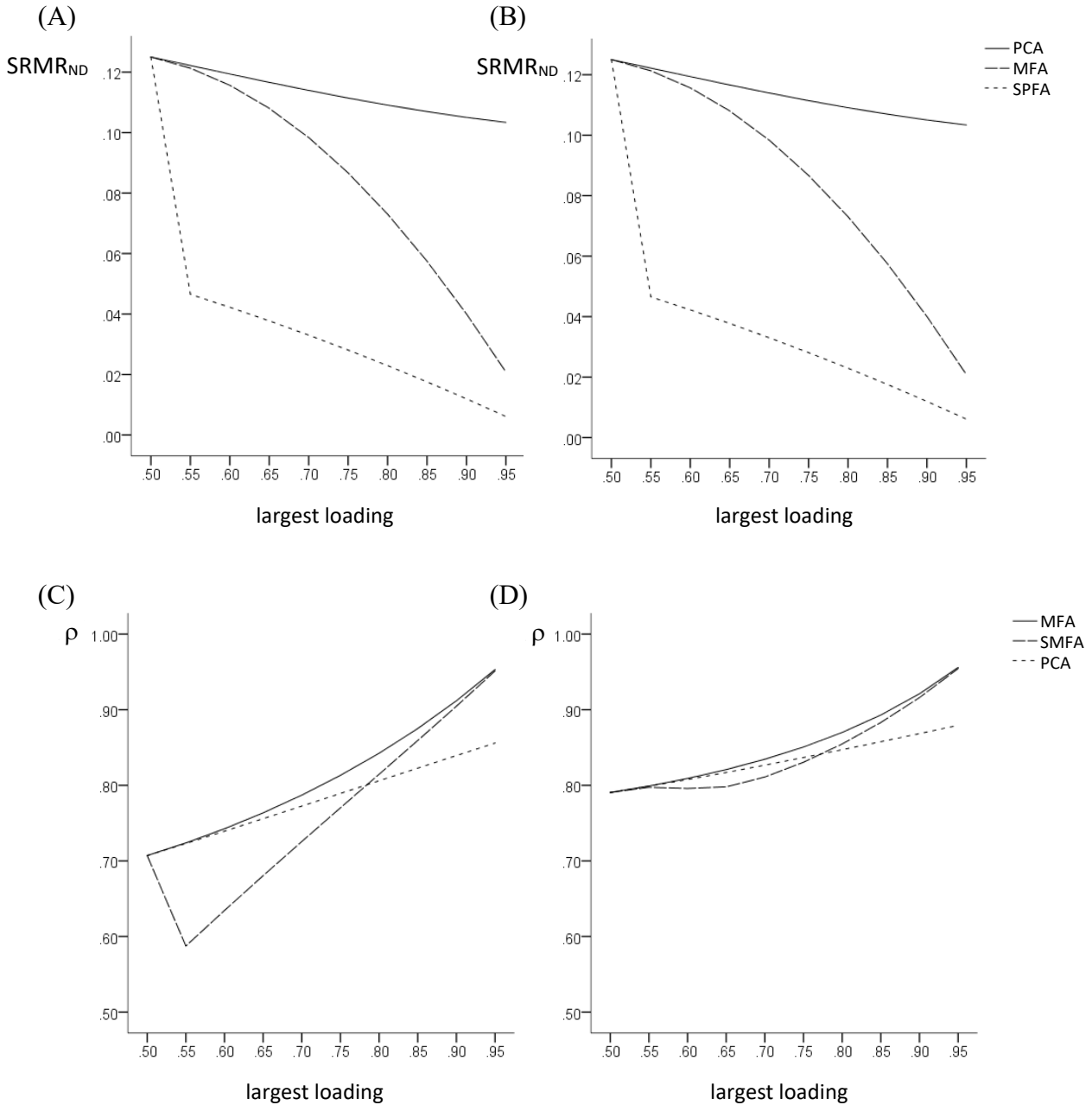

**Figure 1.** Population factor models without model error: SRMR$_{ND}$ based on PCA-, MFA-, and SPFA-scores with (A) $q = 3$, $p = 9$, $p/q = 3$ and (B) $q = 3$, $p = 15$, $p/q = 5$; $\rho$ for PCA-, MFA-, and SPFA-scores with (C) $q = 3$, $p = 9$, $p/q = 3$ and (D) $q = 3$, $p = 15$, and $p/q = 5$

With increasing largest loadings, the SRMR$_{ND}$ of the model implied by MFA factor score predictors becomes more similar to the SRMR$_{ND}$ of the SPFA factor score predictors.

For *p/q* = 5 (Figure 1 B) the differences between the SRMR$_{ND}$ of MFA, SPFA, and PCA are smaller than for *p/q* = 3 (Figure 1 A). It should be noted that the SRMR$_{ND}$ based on SPFA loadings is identical to the SRMR$_{ND}$ based on the SPFA factor score predictor (Theorem 1) and that the SRMR$_{ND}$ based on PCA loadings is identical to the SRMR$_{ND}$ based on component scores (Theorem 2). Since no model error according to the factor model was introduced, the SRMR$_{ND}$ based on MFA loadings was always zero.

The determinacy coefficient, i.e., the correlation ρ of the regression (best linear) factor score predictor (Grice, 2001) based on MFA and on SPFA and of the PCA scores with the factors for the corresponding models are given in Figure 1 C and D. The coefficient of determinacy should regularly be computed in the context of factor analysis (e.g., Grice, 2001; Lorenzo-Seva, & Ferrando, 2013). Moreover, the correlation of the PCA scores with the common factors as well as the correlation of the SPFA factor score predictors with the common factors were computed in order to ascertain how well these scores can be used in order to represent the common factors. Since the correlation of the factor score predictors with the common factor is typically not perfect, it might be possible that PCA scores correlate in a similar magnitude with the common factors as the factor score predictors. Similarly, the SPFA factor score predictor might also correlate with the common factors in a similar magnitude as the factor score predictor. Since the scores of the wanted components are given by $(\mathbf{A'A})^{-1}\mathbf{A'x}=\mathbf{c}$ with $\mathbf{cc'}=\mathbf{I}$, their correlation of with the common factors is

$$\rho_{PCA} = \text{diag}(E(\mathbf{fc'})) = \text{diag}(E(\mathbf{fx'})\mathbf{A}(\mathbf{A'A})^{-1}) = \text{diag}(\mathbf{\Phi\Lambda'A}(\mathbf{A'A})^{-1}). \tag{19}$$

The best linear SPFA factor score predictor is given by $\hat{\mathbf{f}}_{SPFA} = \mathbf{\Lambda'_s\Sigma^{-1}x}$ and the standard deviation of this factor score predictor is $\text{diag}(\hat{\mathbf{f}}_{SPFA}\hat{\mathbf{f}}'_{SPFA})^{1/2} = \text{diag}(\mathbf{\Lambda'_s\Sigma^{-1}\Lambda_s})^{1/2}$ so that its correlation with the factor is given by

$$\rho_{SPFA} = \text{diag}(E(\mathbf{f\hat{f}'})) = \text{diag}(E(\mathbf{fx'})\mathbf{\Sigma^{-1}\Lambda_s})$$
$$= \text{diag}(\mathbf{\Phi\Lambda'\Sigma^{-1}\Lambda_s}\text{diag}(\mathbf{\Lambda'_s\Sigma^{-1}\Lambda_s})^{-1/2}). \tag{20}$$

The results in Figure 1 C and D are based on covariance matrices without error of the factor model. Accordingly, the correlation of the MFA factor score predictors with the factors is always larger than the correlation of the SPFA factor score predictors and the PCA scores with the factor scores. For largest loadings below .70, the PCA scores have larger correlations with the factors than the SPFA factor score predictor, for largest loadings greater .75 the SPFA factor score predictor has larger correlations with the factor than the PCA scores.

**Population simulation for MFA, SPFA, and PCA for data with factor model error**

When model error was introduced according to Tucker, the population correlation matrices were generated from the loadings of major factors and from the loadings of 50 "minor factors" as well as from the corresponding uniqueness. Minor factors have been introduced by Tucker, Koopman, and Linn (1969) in order to represent very small parts of the common variance, which are not part of a given population model. These minor factors therefore represent the "many minor influences" (Tucker et al., 1969), which are not part of the model but that affect the observed scores in the real world. According to MacCallum and Tucker (1991), the loading matrices of 50 minor factors were generated from z-standardized normally distributed random numbers. In the population with large model error, the relative contribution of factors was successively reduced by the factor .95, and the amount of variance explained by the minor factors was set to 30% of the observed total variance. The random numbers were generated by means of the Mersenne Twister random number generator in IBM SPSS 24. All loadings of minor factors were between –1 and +1. Since minor factors were introduced into the population data and since they represent 30% of the total variance, a factor model that is only based on the population loadings of the major factors contains necessarily a substantial amount of model error. In the population with moderate model error, the relative contribution of factors was successively reduced by the factor .85, and the amount of variance explained by the minor factors was set to 20% of the observed total variance.

Orthogonal Procrustes-rotation (Schönemann, 1966) of MFA-, SPFA-, and PCA loadings towards the initial loadings of the major factors was performed in the following in order to assure that different similarities loadings of MFA, SPFA, and PCA to the initial loadings of the major factors are not due to different rotational positions of the factors/components. However, even when the loadings of the three major MFA and SPFA factors were rotated by means of orthogonal Procrustes-rotation towards the initial loading matrix containing only non-zero loadings of .50, the resulting MFA and SPFA loadings are quite different from the initial loadings of the major factors. As an example, Table 1 contains the loadings for the population with large model error. It is remarkable that the maximal loadings of the MFA factors are considerably smaller than the initial maximal loading of .95, even when the MFA loadings are rotated towards the initial loading matrix by means of orthogonal Procrustes-rotation. In contrast, the maximal loadings of the Procrustes-rotated major SPFA factors are close to .95.

**Table 1. Major MFA- and SPFA-factor loadings based on model error and Procrustes-rotation towards the initial (model error free) loadings of the major factors**

| Equal non-zero initial loadings of .50 | | | | | | Non-zero initial loadings of .50 and one non-zero initial loading of .95 | | | | | |
|---|---|---|---|---|---|---|---|---|---|---|---|
| MFA | | | SPFA | | | MFA | | | SPFA | | |
| F1 | F2 | F3 | F1 | F2 | F3 | F1 | F2 | F3 | F1 | F2 | F3 |
| **.46** | -.05 | -.01 | **.40** | -.03 | -.03 | **.47** | -.04 | .03 | **.47** | -.05 | .03 |
| **.62** | -.03 | -.03 | **.98** | -.02 | -.01 | **.53** | -.04 | -.05 | **.51** | -.04 | -.06 |
| **.42** | .01 | .03 | **.38** | -.01 | .03 | **.43** | .02 | -.01 | **.43** | .02 | -.03 |
| **.48** | .03 | -.09 | **.42** | .03 | -.09 | **.49** | .02 | -.07 | **.49** | .02 | -.07 |
| **.32** | .01 | .00 | **.26** | .03 | -.01 | **.73** | .01 | .02 | **.96** | .02 | .01 |
| -.02 | **.50** | -.06 | -.02 | **.52** | -.04 | .01 | **.48** | -.04 | .02 | **.48** | -.04 |
| .01 | **.36** | .04 | .02 | **.34** | .05 | -.03 | **.43** | .03 | -.04 | **.43** | .03 |
| -.01 | **.56** | -.12 | -.01 | **.92** | -.12 | .02 | **.52** | -.07 | .03 | **.52** | -.07 |
| .04 | **.45** | -.07 | .04 | **.46** | -.06 | .04 | **.46** | -.08 | .04 | **.46** | -.09 |
| -.04 | **.50** | .02 | -.03 | **.47** | .05 | -.04 | **.87** | .00 | -.04 | **.98** | -.01 |
| -.07 | -.03 | **.50** | -.09 | -.01 | **.50** | -.02 | -.03 | **.53** | -.01 | -.05 | **.56** |
| -.03 | -.07 | **.58** | -.02 | -.06 | **.90** | -.03 | -.05 | **.54** | -.04 | -.05 | **.56** |
| .04 | -.06 | **.46** | .04 | -.04 | **.47** | .04 | -.07 | **.48** | .04 | -.08 | **.50** |
| -.02 | -.01 | **.52** | -.01 | -.02 | **.58** | -.06 | .01 | **.50** | -.07 | .00 | **.51** |
| -.02 | -.03 | **.39** | -.04 | .01 | **.39** | -.01 | -.01 | **.76** | -.02 | -.02 | **.94** |

*Note.* Main loadings were given in bold face.

The model error of the MFA-, SPFA-, and PCA-models that were based on three major factors with initial maximal non-zero loadings of .50 to .95 are reported in terms of the $SRMR_{ND}$ (see Figure 2 A and C). Even in data that were based on the model error of the factor model, the fit of the MFA loadings was the best. However, the $SRMR_{ND}$ of the models implied by the factor score predictors or component scores was smaller for the SPFA- than for the MFA- and PCA-models.

The correlation ρ of the regression (best linear) factor score predictor based on MFA and on SPFA and of the PCA scores with the major factors for the corresponding models based on large model error are given in Figure 2 B. For largest initial major factor loadings below .75, the PCA scores have larger correlations with the factors than the MFA and SPFA factor score predictors. For largest initial factor loadings greater .75 the SPFA factor score predictor has larger correlations with the factor than the MFA factor score predictor and the PCA scores. Thus, in presence of large model error of the factor model, the SPFA factor score predictor can be a more valid measure of the original major factor than the corresponding MFA factor score predictor. This occurs probably in presence of a model error leading to underestimation of the initial maximal loadings on the major factors by means of MFA (Table 1). Under such conditions, SPFA may result in a more accurate estimation of the initial maximal loadings and thereby in a more valid factor score predictor. It should also be noted that the mean determinacy coefficient was slightly larger for PCA scores than for MFA factor score predictor when the size and variability of the main loadings small (see Figure 2 B). Thus, for the respective population data, the maximum determinacy coefficient was obtained for PCA for small variability of main loadings and for SPFA for a larger variability of main loadings. When model error was moderate, the determinacy coefficient of the MFA factor score predictor was mostly larger than the determinacy coefficient of the SPFA factor score predictor (Figure 2 D).

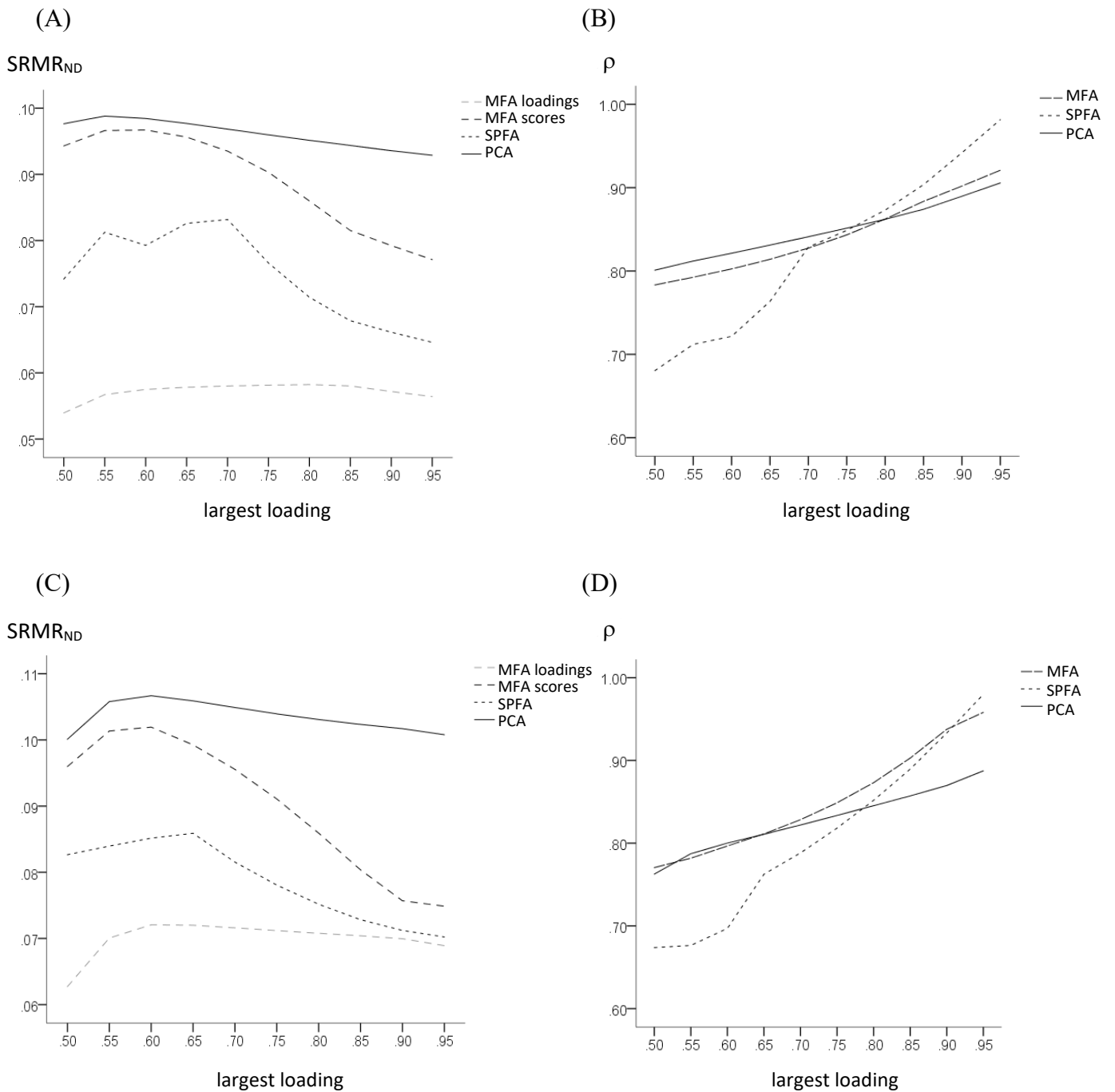

**Figure 2.** Population factor models with $q = 3$, $p = 15$, $p/q = 5$; large model error: (A) $SRMR_{ND}$ for MFA loadings, MFA-scores, PCA, and SPFA; (B) $\rho$ for PCA-, MFA-, and SPFA-scores; moderate model error: (C) $SRMR_{ND}$ for MFA loadings, MFA-scores, PCA, and SPFA; (D) $\rho$ for PCA-, MFA-, and SPFA-scores

**Sample simulation for MFA, SPFA, and PCA**

A sample simulation was performed in order to investigate the size of the $SRMR_{ND}$ and the determinacy coefficient ( $\rho$ ) for MFA, SPFA, and PCA when sampling error occurs across a number of different conditions. Again, the Mersenne Twister random number generator was used for the generation of random numbers. The simulation was performed for $q = 3$ factors with $p = 15$ variables and for $q = 6$ with $p = 30$, for n = 150 and 600 cases, for $p/q = 5$ variables with non-zero population loadings per factor, and for non-zero initial population loadings $\Lambda_i$ = 35, .40, .45, and .50. At each $\Lambda_i$-level, one of the non-zero population loadings per factor was increased to a maximum loading of the model $\Lambda_m$ by .00, .15, .30, and .45. For example, for the $\Lambda_i$ = .40 solutions, there was one simulation where all non-zero loadings were $\Lambda_i$ = .40, one simulation with one loading $\Lambda_m$ = .55 per factor, one simulation with $\Lambda_m$ = .70 per factor, and one simulation with $\Lambda_m$ = .95 per factor. The simulations were performed with model error and without model error. Model error was based on 50 minor factors, which were generated from z-standardized normally distributed random numbers. A moderate amount of model error was introduced, where the relative contribution of factors was successively reduced by the factor .85, and the amount of variance explained by the minor factors was set to 20% of the observed total variance. For each level of $\Lambda_i$ and $\Lambda_m$ the same set of minor factors was used in order to introduce the same population model error across conditions. With a constant set of minor factors, changes in the $SRMR_{ND}$ and $\rho$ can clearly be attributed to $\Lambda_i$ and $\Lambda_m$ and not to a changing set of minor factors. Overall, there were 128 conditions, i.e., 16 (= 4 $\Lambda_i$-levels × 4 $\Lambda_m$-levels) loading patterns × 2 numbers of factors × 2 sample sizes × 2 levels of model error. For each of the 128 population conditions 1,000 samples were drawn and analyzed by means of MFA, SPFA, and PCA. As for the population simulation, orthogonal Procrustes-rotation (Schönemann, 1966) towards the initial population loadings of the major factors was performed in order to assure that different similarities of the MFA-, SPFA-, and PCA loadings to the initial loadings of the major factors are due to the method of variance extraction and not to different rotational positions of the factors/components. For a subset of the sample based simulation conditions ($q = 6$, $\Lambda_i$ = .35 and .50, all $\Lambda_m$-levels, n = 150 and 600, without and with model error), Varimax-rotation was performed by means of the gradient-projection algorithm provided by Jennrich (2001) and Bernaards and Jennrich (2005) in order to compare effects of factor rotation on $\rho$ for MFA, SPFA, and PCA.

The results of the sample based simulation study are as follows: For the simulation without model error, the mean SRMR$_{ND}$ was smallest when based on the loadings of MFA model, was larger when based on the SPFA scores, again larger when based on the MFA scores, and largest when based on PCA scores (see Figure 3). This result corresponds to Equation 17. It was not necessary to show the SRMR$_{ND}$ for the loadings based on the SPFA and PCA because it is equal to the SRMR$_{ND}$ based on the respective scores (Theorem 1 and 2). The SRMR$_{ND}$ decreases with increasing $\mathbf{\Lambda}_m$ so that the SRMR$_{ND}$ become more similar for the loadings of the MFA model, the SPFA scores, and the MFA scores. The effect of sample size on mean SRMR$_{ND}$ is not very important, given that the larger sample was four times larger than the smaller sample. However, the standard deviation of the SRMR$_{ND}$ decreased with increasing sample size and with a larger number of factors. Moreover, a larger number of factors resulted in a smaller SRMR$_{ND}$ differences between MFA model, SPFA scores, MFA scores, and PCA scores (see Figure 3). Results were rather similar for the condition with model error (see Figure 4). The only substantial difference of this condition is that the effect of $\mathbf{\Lambda}_m$ on the SRMR$_{ND}$ was less pronounced. This implies that the differences between the SRMR$_{ND}$ of MFA and SPFA do not decrease substantially when $\mathbf{\Lambda}_m$ increases.

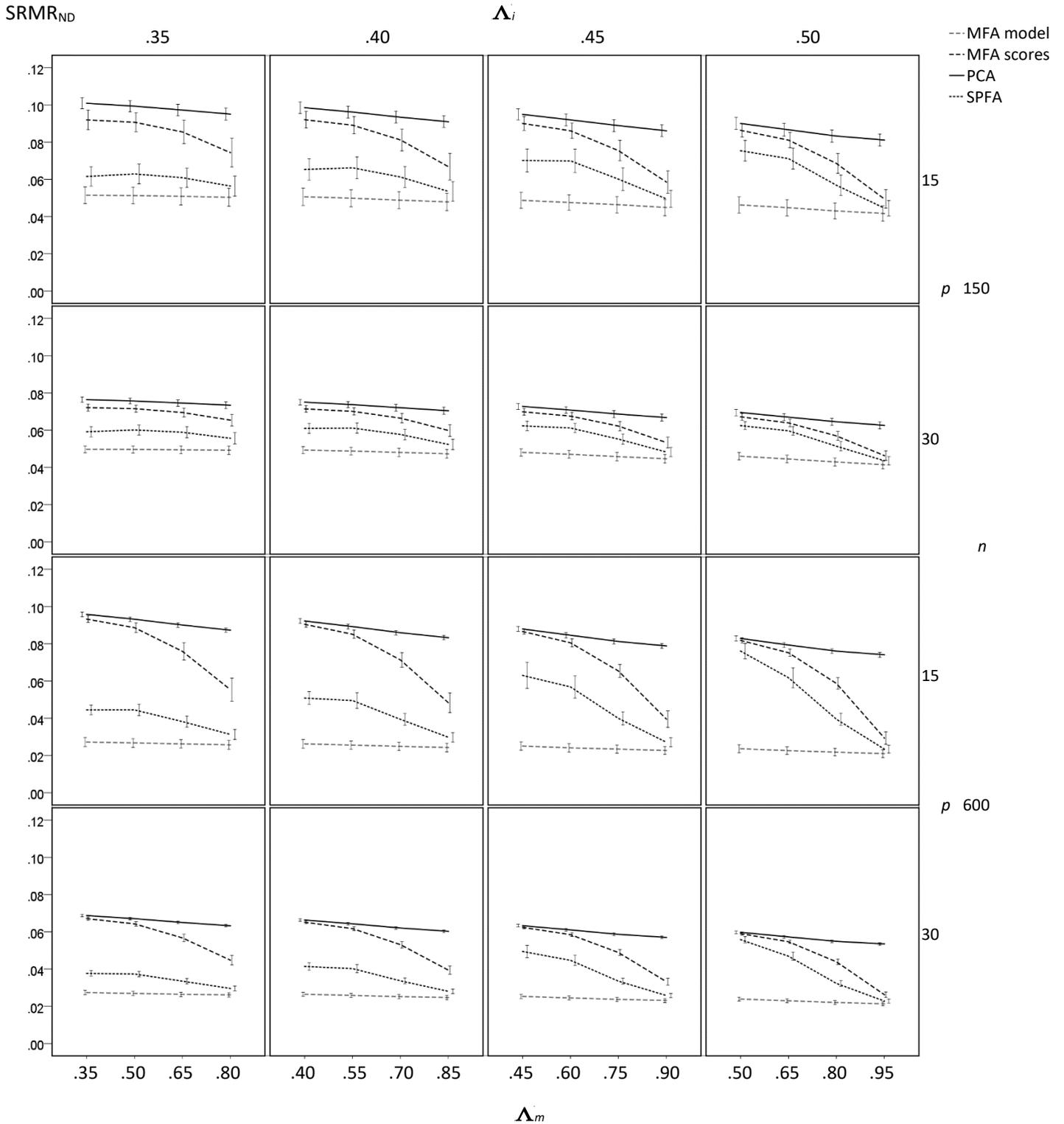

**Figure 3.** Mean SRMR$_{ND}$ based on samples drawn from population factor models without model error for $q = 3$ with $p = 15$ and $q = 6$ with $p = 30$ for the MFA-model, MFA-scores, PCA-scores, and SPFA-scores; $\Lambda_i$ = initial population loadings; $\Lambda_m$ = maximal population loadings; the error bars represent standard deviations

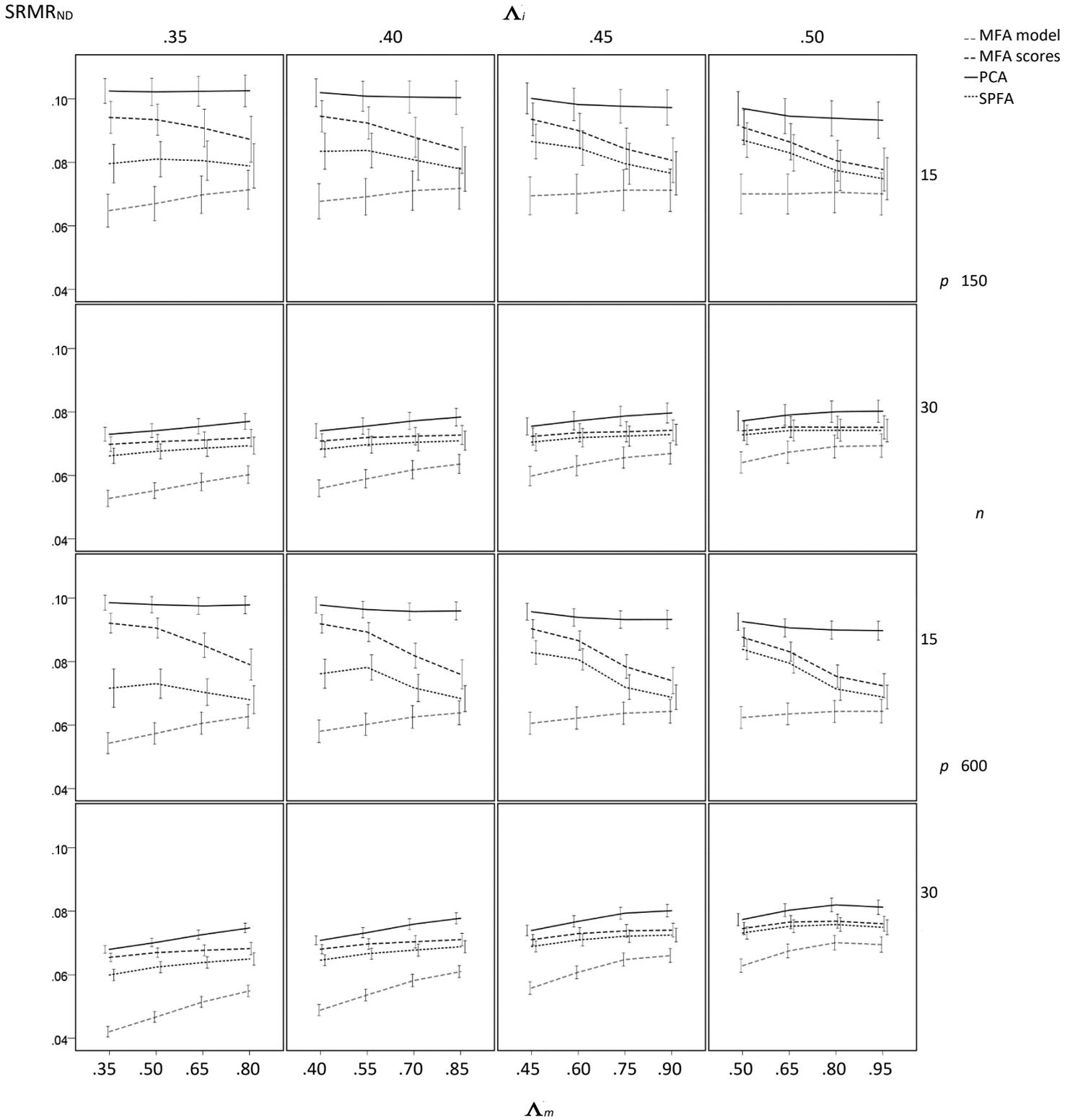

**Figure 4.** Mean SRMR$_{ND}$ based on samples drawn from population factor models with model error for $q = 3$ with $p = 15$ and $q = 6$ with $p = 30$ for the MFA-model, MFA-scores, PCA-scores, and SPFA-scores; $\Lambda_i$ = initial population loadings; $\Lambda_m$ = maximal population loadings; the error bars represent standard deviations

In the condition without model error, the mean determinacy coefficient ($\rho$) was largest for PCA when $\Lambda_m$ was small, it was largest for MFA when $\Lambda_m$ was large, and it was smallest for SPFA when $\Lambda_m$ was small (see Figure 5). When $\Lambda_m > .75$ mean $\rho$ was larger for SPFA than for PCA and comes close to MFA. For increasing overall level of salient loadings ($\Lambda_i$), the mean $\rho$ becomes more similar for MFA, SPFA, and PCA. Moreover, the standard deviation of $\rho$ was considerably smaller for the larger sample size. In the condition with model error, the mean $\rho$ was smallest for SPFA only for models with the smaller number of factors, small $\Lambda_i$, and small $\Lambda_m$. For the larger number of factors, small $\Lambda_i$, and large $\Lambda_m$, the mean $\rho$ was slightly larger for SPFA than for MFA and PCA (see Figure 6). Again, for larger $\Lambda_i$ the mean $\rho$ becomes more similar for MFA, SPFA, and PCA. The effect of sample size on the standard deviations was less pronounced in the condition with model error than in the condition without model error.

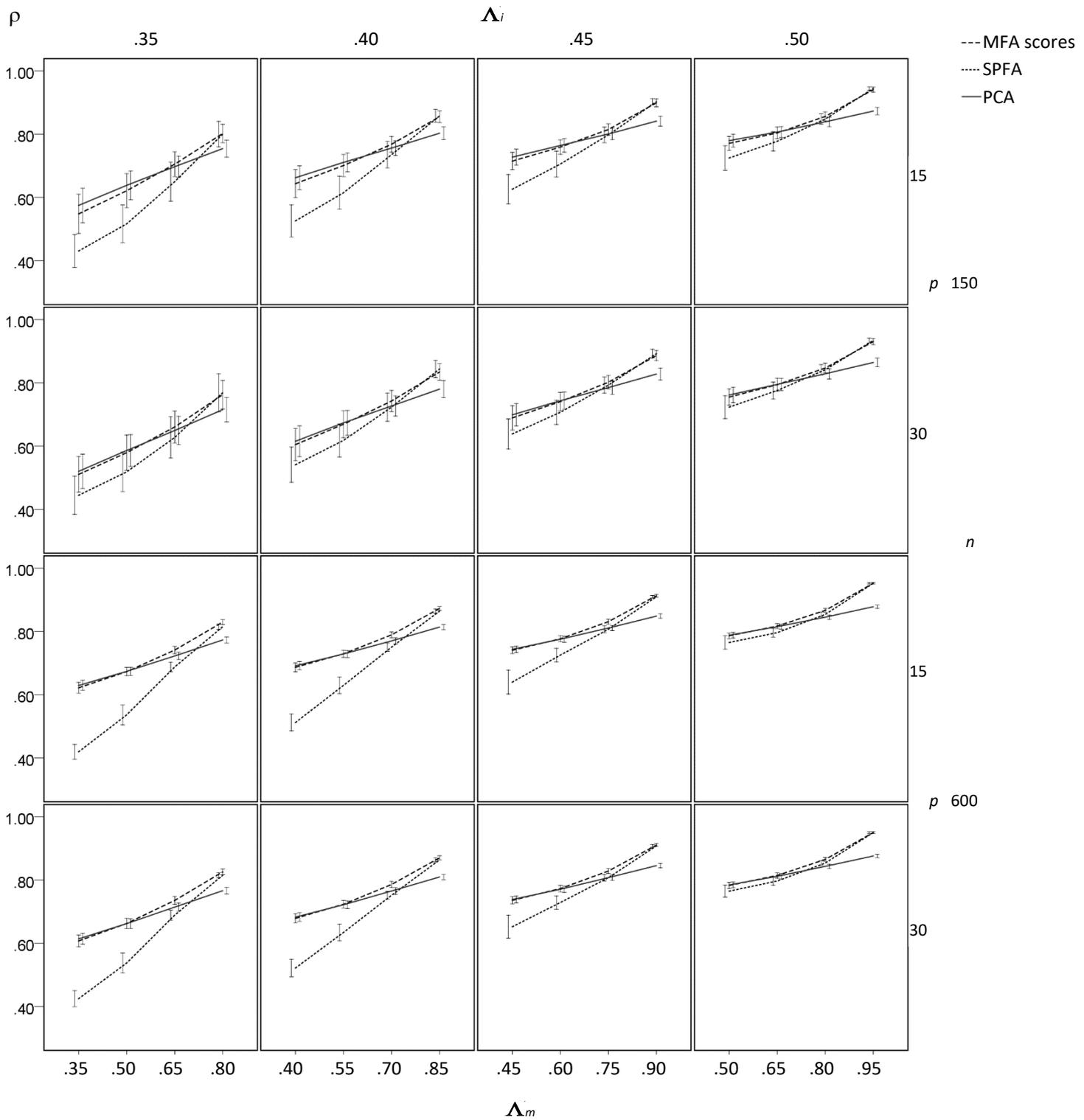

**Figure 5.** Mean determinacy coefficient ( $\rho$ ) based on samples drawn from population factor models without model error for $q = 3$ with $p = 15$ and $q = 6$ with $p = 30$ for the MFA-scores, PCA-scores, and SPFA-scores; $\Lambda_i$ = initial population loadings; $\Lambda_m$ = maximal population loadings; the error bars represent standard deviations

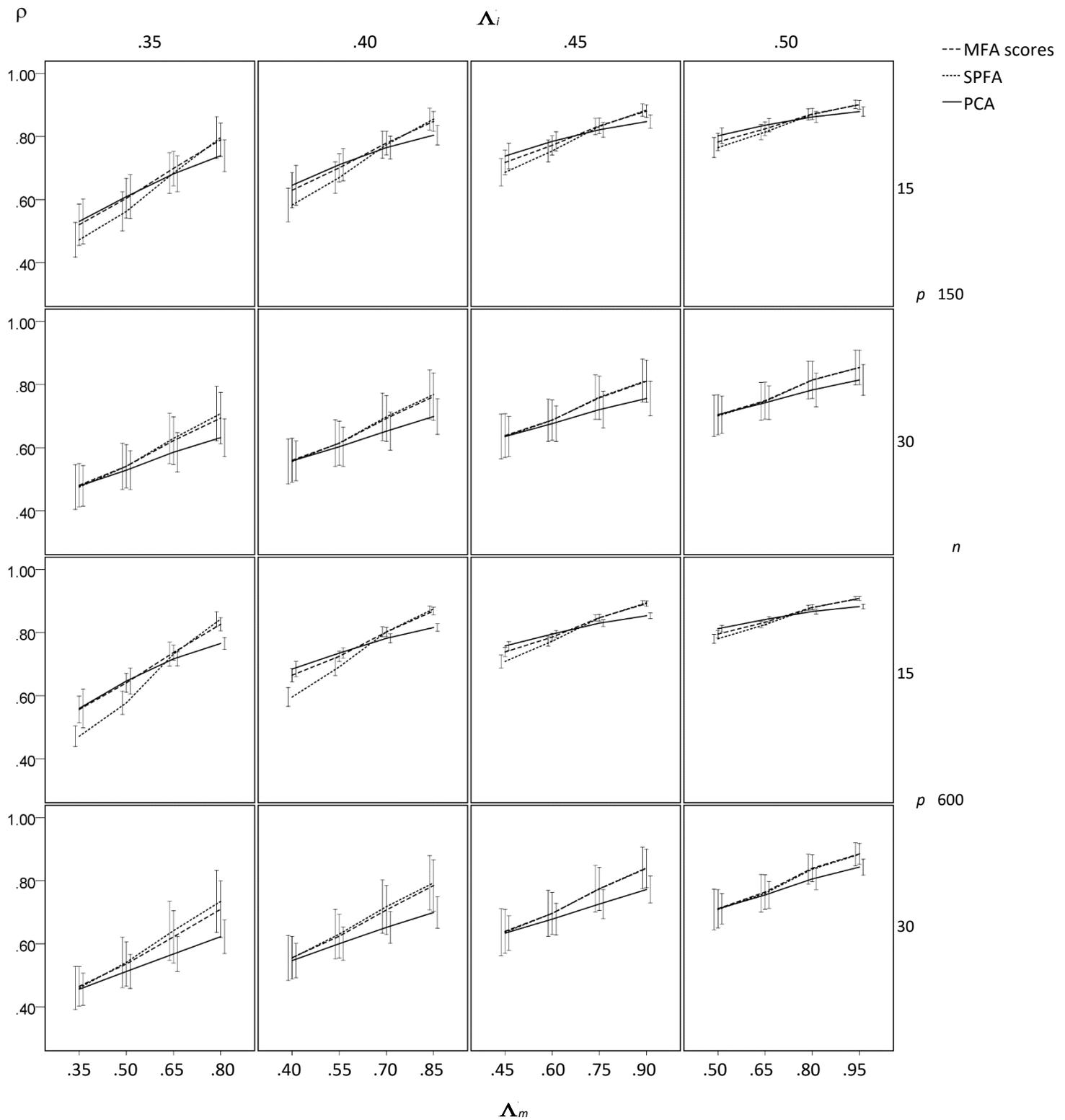

**Figure 6.** Mean determinacy coefficient ($\rho$) based on samples drawn from population factor models with model error for $q = 3$ with $p = 15$ and $q = 6$ with $p = 30$ for the MFA-scores, PCA-scores, and SPFA-scores; $\Lambda_i^{'}$ = initial population loadings; $\Lambda_m^{'}$ = maximal population loadings; the error bars represent standard deviations

For the conditions with q = 6, for n = 150 and n = 600, with and without model error, the mean ρ after Varimax-rotation was also computed for MFA, SPFA, and PCA (Figure 7). Overall, the differences were small and the mean ρ for SPFA reached the level of MFA. No substantial advantages occurred for PCA when Varimax-rotation was performed.

**Discussion**

The starting point of the present study was that the model implied by the factor score predictors of the common factor model does not reproduce the non-diagonal elements of the observed covariance matrix as well as the common factors do (Beauducel & Hilger, 2015). In order to close this gap, it was proposed to estimate factor loadings in a way that not the loadings but the factor score predictors computed from the loadings optimally reproduce the non-diagonal elements of the observed covariance matrix. This estimation method is termed 'Score Predictor Factor Analysis' (SPFA) and based on a similar estimation procedure as MFA, with the difference that the estimation is not initially based on the loadings, but on the model implied by the SPFA factor score predictors. It is shown that the SPFA loadings reproduce the same non-diagonal elements of the observed covariance matrix as the model implied by the SPFA factor score predictors (Theorem 1). It was also shown algebraically that the SPFA loadings will reproduce the non-diagonal elements of the observed covariances less or equally well than the MFA loadings and that the model implied by the SPFA score predictor will reproduce the non-diagonal elements of the observed covariances equally or better than the model implied by the MFA factor score predictor (Equation 17). Therefore, when the focus is on the loadings, there are reasons for preferring a factor model like MFA and when the focus is on the factor score predictor, there are reasons for preferring SPFA. Moreover, it is an empirical question whether the error of the factor model (MacCallum, 2003) or the error of SPFA is more substantial.

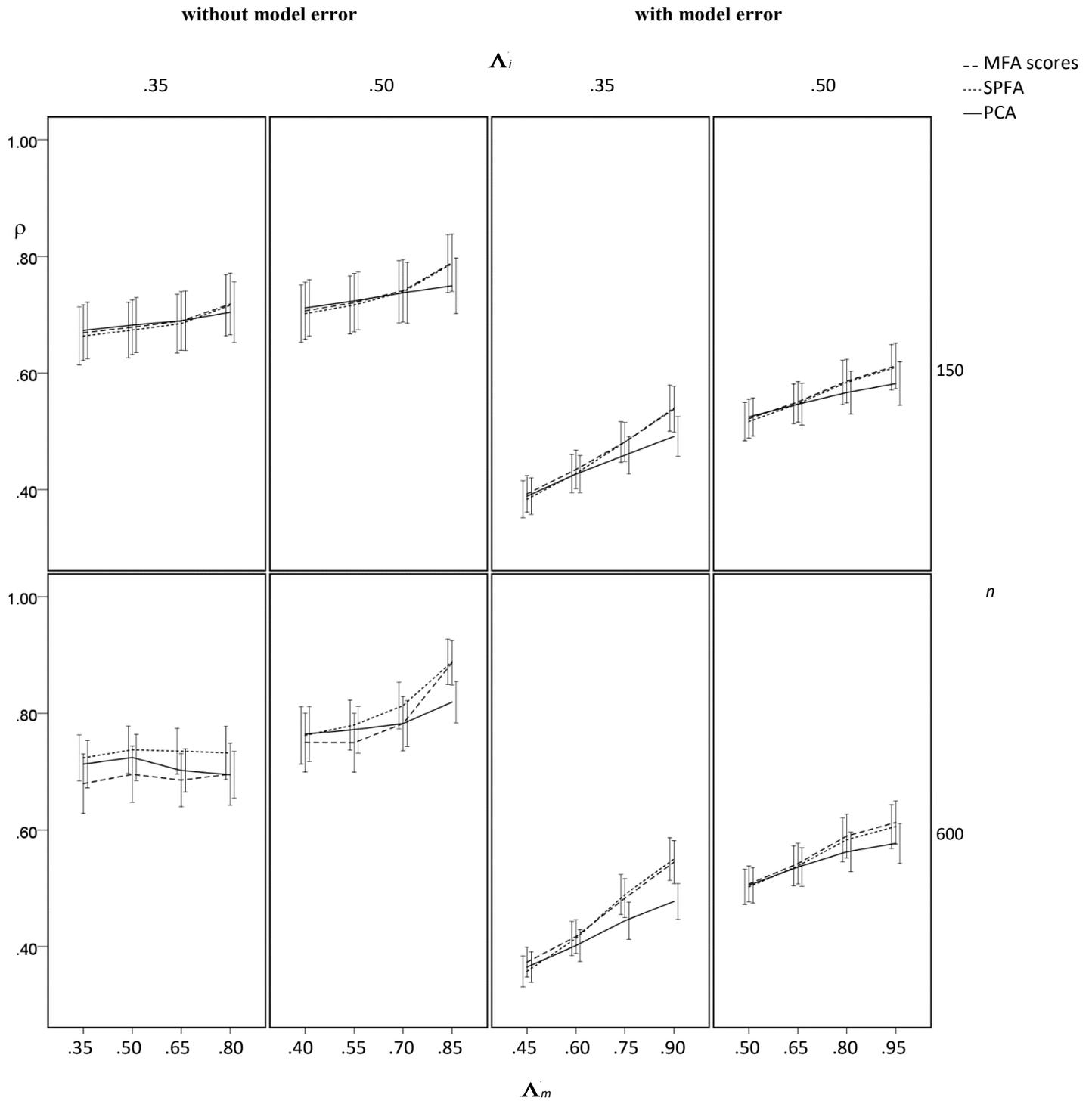

**Figure 7.** Mean determinacy coefficient ( $\rho$ ) based on samples drawn from population factor models with model error for $q = 6$ with $p = 30$ for the MFA-scores, PCA-scores, and SPFA-scores; $\Lambda_i$ = initial population loadings; $\Lambda_m$ = maximal population loadings; the error bars represent standard deviations

The evaluation of the SPFA was based on a comparison with MFA and PCA by means of a population based simulation study and a sample based simulation study. One dependent variable of the simulation studies was the standardized root mean square residual based on the non-diagonal elements of the observed covariance matrix ($SRMR_{ND}$). This fit index allows to investigate how well the intended goal of SPFA to reproduce the non-diagonal elements of the observed covariance matrix by means of the model implied by the SPFA factor score predictors can be achieved. Moreover, the mean coefficient of determinacy ( $\rho$ ) was computed in order to investigate the correlation of the MFA-, SPFA-best linear factor score predictors, and PCA-components with the population factors.

It turns out that the $SRMR_{ND}$ based on the SPFA factor score predictor was consistently smaller in the simulations based on the populations and the samples than the $SRMR_{ND}$ based on the MFA factor score predictor and PCA scores. The $SRMR_{ND}$ based on PCA scores was consistently larger than the $SRMR_{ND}$ based on the MFA- and SPFA-factor score predictor. These results were also found when the simulations were based on population data that do not fit perfectly to the factor model (i.e. when model error occurs). Thus, the model implied by the SPFA-factor score predictor allows for an optimal score-based reproduction of the non-diagonal elements of the observed covariance matrix. Therefore, SPFA-factor score predictors may be of interest when factor score predictors are to be computed and when they should represent a model that optimally reproduces the non-diagonal elements of the observed covariance matrix.

Finally, when there was some population error of the factor model and when there were a few large main loadings, the SPFA factor score predictor had a larger coefficient of determinacy than the MFA factor score predictor. Thus, in case of bad model fit and when a few large main loadings occur, the SPFA factor score predictor may be chosen in order to represent the population factors most appropriately. Moreover, when there were rather small and nearly equal main loadings, the PCA score had a larger coefficient of determinacy than the MFA factor score predictor and than the SPFA factor score predictor, so that the PCA score may represent the original factors most appropriately under these conditions. Although the differences between PCA and the factor model have often been regarded as negligible (Velicer & Jackson, 1990), differences between these models could occur from the perspective of individual scores. One strategy to decide whether the MFA factor score predictor, the SPFA factor score predictor, or PCA scores should be computed, could be to chose the model with the largest coefficient of determinacy. The Equations 19 and 20 that were given for the computation of determinacy coefficients for PCA scores and SPFA factor score predictors can be used for this purpose. If the SPFA factor score predictor or the PCA score outperforms the MFA factor score predictor in terms of the correlation with the MFA factor, this could be a good reason to chose one of these alternative models.

However, whereas the results on the $SRMR_{ND}$ are not affected by factor rotation, the results for the determinacy coefficient depend on factor rotation. The effects of factor rotation on results were eliminated in the first simulation study by means of orthogonal Procrustes rotation in order to compare MFA, SPFA and PCA as methods for the extraction of variance. In a last simulation, congruence coefficients for MFA, SPFA and PCA were compared for Varimax-rotated factors. It turns out that the differences between the methods were less marked when Varimax-rotation was performed. There was no substantial decrease of the congruence coefficients when SPFA was combined with Varimax-rotation. Thus, SPFA-factor score predictors optimally reproduce the non-diagonal elements of the observed covariance matrix and provide similar congruence coefficients as MFA. Therefore, SPFA estimation of factor loadings might be of interest when the factor score predictors are important.

Further research should be performed in order to investigate SPFA estimation of loadings in further methods of factor rotation, in the context of larger numbers of factors and variables. The precision of methods that allow to determine the number of factors to extract should also be investigated in the context of SPFA.